\documentclass[dvips]{acta}
\usepackage{supertabular,lscape,epsfig}
\usepackage{amssymb}
\usepackage{amsmath}
\usepackage{multirow}

\begin{document}

\begin{Titlepage}
\Title{2009 Superoutburst of Dwarf Nova 1RXS J053234.9+624755}
\Author{A.~Rutkowski$^1$, A.~Olech$^2$, R.~Poleski$^{3}$, M.~Sobolewska$^{4,5,6}$
P.~Kankiewicz$^{7}$,T.~Ak$^{1,8}$ and D.~Boyd$^9$}
{{$^1$TUBITAK National Observatory, Akdeniz University Campus, 07058 Antalya, Turkey\\
e-mail:artur@tug.tug.tubitak.gov.tr\\}
{$^2$Nicolaus Copernicus Astronomical Center, Polish Academy of Sciences,\\
Ul. Bartycka 18, 00-716 Warszawa, Poland\\ }
{$^{3}$ Warsaw University Observatory, Al. Ujazdowskie 4, 00-478 Warsaw,
Poland \\ }
{$^4$ Smithsonian Astrophysical Observatory, 60 Garden Street,
  Cambridge, MA 02138, USA\\}
{$^5$ Foundation for Research and Technology - Hellas, IESL, Voutes, 71110
  Heraklion, Crete, Greece\\} 
{$^6$ University of Crete, Department of Physics, Voutes, 71003 Heraklion,
  Crete, Greece\\} 
{$^7$ Astrophysics Division, Institute of Physics, Jan Kochanowski University, Swietokrzyska~15, 25-406 Kielce, Poland \\}
{$^8$Istanbul University, Faculty of Sciences, Department of Astronomy and Space Sciences, 34119 University, Istanbul, Turkey \\ }
{$^9$British Astronomical Association, Variable Star Section, West Challow OX12 9TX, England }
}
\Received{Month Day, Year}
\end{Titlepage}
\Abstract{We present photometric observations of the dwarf nova 1RXS
J053234.9+624755. We performed a detailed analysis of the superoutburst
that occurred in August 2009. 
We found the superhump period to be $P_{sh}=0.057122(14)$days.
Based on the $O-C$ diagram we conclude that $P_{sh}$ increased
during the plateau at the rate of $dP_{sh}/dt=(9.24 \pm 1.4) \cdot 10^{-5}$.
Both the $O-C$ analysis and evolution of the superhumps light curve favour
the model in which superhumps originate in a variable source 
located in the vicinity of the hot spot.
In addition, the evolution of the light curve suggests that the superhump light 
source approaches the disc plane as the superoutburst declines.
Detailed analysis of the superoutburst plateau phase enabled us to detect a signal which
we interpret as apsidal motion of the accretion disc.
We detected additional modulations during the final stage of
the superoutburst characterized by periods of 104s and 188s which we tentatively
interpret as quasi periodic oscillations.
Estimations of $A_{0}$ and $A_{n}$ are in agreement with
the dependence discovered by Smak (2010) between the 
amplitude of superhumps and the orbital inclination.
}

{accretion, accretion discs - binaries: cataclysmic variables, stars: dwarf
novae, oscillations, stars: individual: 1RXS J053234.9+624755, 2MASS 
J05323386+6247 520,USNO-B1.0 1527-00176070, USNO-B1.0 1527-00176070}
\section{Introduction}
Over 30 years have passed since the discovery of superhumps in cataclysmic variables
(Vogt 1974, Warner 1995). Different physical processes have been
proposed to explain the origin of the superhump phenomenon.
Models considered have included the ejection of matter from the white
dwarf due to pulsation instabilities (Vogt 1974), periodic modulation
of dissipation in the elliptical and precessing accretion disc (Osaki 1989),
or models which explain superhumps in terms of the oscillations of 
a hot spot caused by uneven stream flows from the secondary (Smak 2009c).
Nowadays we posses enough knowledge about the components of dwarf
novae to describe these systems completely. 
It seems likely that under the strength of argument provided by Smak (2010),
most doubts regarding superhumps will be dispelled soon.
However, each new model should be tested by observations
and therefore it is important to observe dwarf novae especially 
during superoutbursts, i.e. when superhumps occur predominantly.

1RXS J053234.9+624755 (1RXS J0532) is an example of a dwarf nova with documented superoutbursts.
It was discovered quite unexpectedly as a counterpart of an X-ray source in the ROSAT
all-sky bright catalog by Bernhard et al. (2005).
Detailed investigation of the Sonneberg Plate Archive for this object
found eight outbursts between 1990 and 2005 with a mean interval of 133.6d.
Soon, further observations resulted in the detection of superhumps in the light curve of
1RXS J0532 and revealed that this star is a SU UMa type dwarf nova (Poyner \& Shears 2006). Monitoring of
superoutbursts showed the evolution from a clear tooth-shape
light curve variation to more random flickering (Parimucha \& Dubovsky 2006).
Based on spectroscopic observations an orbital period of 0.05620(4) d was reported by Kapusta \& Thorstensen (2006). 
An observational campaign during the 2005 superoutburst reported by 
Imada et al. (2009) showed the superoutburst accompanied by a precursor.
The light curve of the precursor revealed a gradual increase in the amplitude
of the light variation which was interpreted as developing superhumps.
As this is in contradiction to the standard Thermal - Tidal Instability (TTI) model,
further interpretation was based on the "refined" TTI model.
The authors concluded that the existence of the precursor and the presence
of growing superhumps during the precursor are related to the mass ratio 
of the system. This in turn determines the 3:1 resonance radius and 
the tidal truncation radius ratio which are supposed to be responsible 
for the behavior of the superhumps (Osaki 2005).
Moreover, the photometric data provided an estimate of the length of the supercycle,
$\sim$450 days, and the superhump period, $P_{sh}=0.57169(6)$d (Imada et al. 2009).
\section{Observations}
Our observational campaign of 1RXS J0532 lasted from 24 August 2009 to 10 September 2009,
and was performed as part of the long-term observational 
project - CURious Variable Experiment (CURVE, Olech et al. 2009,
Rutkowski et al. 2009). In this paper we report observations that cover 17 nights and include 7271 measurements.
Thanks to the CURVE collaboration we were able to use
several telescopes in different locations. This greatly minimized the effect of bad weather
on our campaign, so observational gaps were generally shorter than one day.

\MakeTable{c c c c c c c c c}{12.5cm}{Observational journal of the 2009 1RXS J0532 campaign.}{
\hline \hline
Date in    & Start [HJD]  & End [HJD]    & Dur.  & Obs.    & No. of & $<V>$ & $A_{max}$ \\
2009       & 2455000+     & 2455000+     & [hr]  & index   & points & mag   & mag \\
\hline
Aug 24    & 68.454740     & 68.534170   & 1.90  &  West Ch.& 197    & 11.97 & $0.039\pm0.007$ \\
Aug 25    & 69.342298     & 69.522114   & 4.31  &  Kielce  & 120    & 12.24 & $0.151\pm0.010$ \\
Aug 28    & 72.457581     & 72.634828   & 4.25  & Skinakas & 625    & 11.93 & $0.110\pm0.005$ \\
Aug 29    & 73.463275     & 73.626979   & 3.93  & Skinakas & 450    & 12.10 & $0.114\pm0.005$ \\
Aug 30    & 74.449727     & 74.548565   & 2.37  & Skinakas & 247    & 12.32 & $0.085\pm0.007$ \\
Aug 31    & 75.484640     & 75.631456   & 3.52  & Skinakas & 822    & 12.39 & $0.078\pm0.015$ \\
Aug 31    & 75.508670     & 75.613440   & 2.51  & Ostrowik & 171    & 12.40 & ---           \\
Sep 01    & 76.327500     & 76.615850   & 6.92  & Ostrowik & 425    & 12.35 & $0.106\pm0.005$ \\
Sep 02    & 77.468199     & 77.632007   & 3.93  & Skinakas & 319    & 12.40 & $0.131\pm0.005$ \\
Sep 02    & 77.322880     & 77.423630   & 2.42  & Ostrowik & 96     & 12.44 & ---           \\
Sep 03    & 78.459024     & 78.633852   & 4.20  & Skinakas & 947    & 12.42 & $0.152\pm0.010$ \\
Sep 04    & 79.463728     & 79.621767   & 3.79  & Skinakas & 974    & 12.57 & $0.146\pm0.010$ \\
Sep 05    & 80.479105     & 80.627843   & 3.57  & Skinakas & 1047   & 13.12 & $0.173\pm0.012$ \\
Sep 05    & 80.441150     & 80.619560   & 4.28  & Ostrowik & 278    & 13.09 & ---          \\
Sep 07    & 82.363320     & 82.477090   & 2.73  & Ostrowik & 154    & 14.50 & $0.55\pm0.05$ \\
Sep 08    & 83.364140     & 83.620240   & 6.14  & Ostrowik & 167    & 14.52 & $0.50\pm0.05$ \\
Sep 09    & 84.367280     & 84.623550   & 6.15  & Ostrowik & 174    & 14.78 & $0.85\pm0.01$ \\
Sep 10    & 85.353360     & 85.536260   & 4.39  & Ostrowik & 46     & 14.40 & $0.81\pm0.05$ \\
\hline
}

The data were collected mainly by two telescopes: the 1.3-m Ritchey - Chretien telescope of the Skinakas Observatory, Crete, Greece, and
the 0.6-m Cassegrain telescope of Warsaw University Observatory. In Skinakas Observatory
we used "white light" and Johnson $B,V,R,I$ filters. We observed with the ANDOR 2048x2048 back-illuminated
CCD array with $13.5 \mu m$ pixels. The Warsaw University Observatory was
equipped with the Tektronix 512x512 TK512CB back-illuminated CCD camera with
$27 \mu m$ pixels.

In addition, two one-night observing runs were carried out by two smaller telescopes.
The first run on 24 August was carried by a 25-cm Newtonian 
telescope equipped with a Starlight Xpress HX-516 CCD and located
in West Challow, England. The second run on 25 August was performed by a 35-cm Schmidt - Cassegrain telescope
equipped with a ST-7 CCD located in Kielce, Poland (Jan Kochanowski University of Humanities and Sciences).

Exposure times varied from 10 sec (for the 1.3-m telescopes) 
to 120 sec (for the 35-cm telescopes). Exposures with the 1.3-m Skinakas telescope
were autoguided. For the other telescopes, even without autoguiding, PSF profiles seemed undistorted and mostly circular.

The IRAF package was used for data reduction and PSF photometry was performed with DAOphotII.
The majority of measurements were made without filters (in white light).
Occasionally, when using the 1.3-m telescopes, $B,V,R,I$ exposures were taken to
calibrate the global light curve. 
Performing the observations in white light allowed us to take shorter exposures and 
minimize guiding errors. However this approach
could potentially introduce uncertainty when comparing measurements from
different photometric systems.

Relative unfiltered magnitudes of 1RXS J0532 were determined as the 
difference between the magnitude of the variable and the magnitude of a nearby
comparison star, 2MASS J05323636+6245536. The SIMBAD database provided the $B,V$ magnitudes for this star.
To find R and I magnitudes we selected another comparison star, 2MASS J05325096+6246161.
We also investigated other possible comparison stars in the field
but these all introduced larger errors and more scatter in the light curve. 

We adopted the following magnitudes for 2MASS J05323636+6245536:
$B=11.49\pm0.01$, $V=11.21\pm0.01$, $R= 10.08\pm0.01$, $I=10.69\pm0.02$.
We assumed that our white light measurements introduced 
constant shifts with respect to standard $V$ magnitudes. 
Twelve $V$ measurements were obtained during the superoutburst and
used to estimate the $V$ magnitude for the rest of the collected data. The light curves
in white light have been shifted by a constant value to agree with
the points obtained in the $V$ filter. We estimate the errors introduced by his procedure  to be smaller than $\sim0.2$mag.

Table 1 presents a journal of our CCD observations of 1RXS J0532. Successive columns
contain the date, start and end of the observational run in HJD; duration of the observation
during a particular night; location of the observatory; number of points gathered during the night;
the approximate average brightness in~$V$; and the maximum superhump amplitude.
In total, 1RXS J0532 was observed for over 73 hours.
\section{Global light curve and its color variation}
\begin{figure}[ht]
\begin{center}
\includegraphics[width=0.8\textwidth]{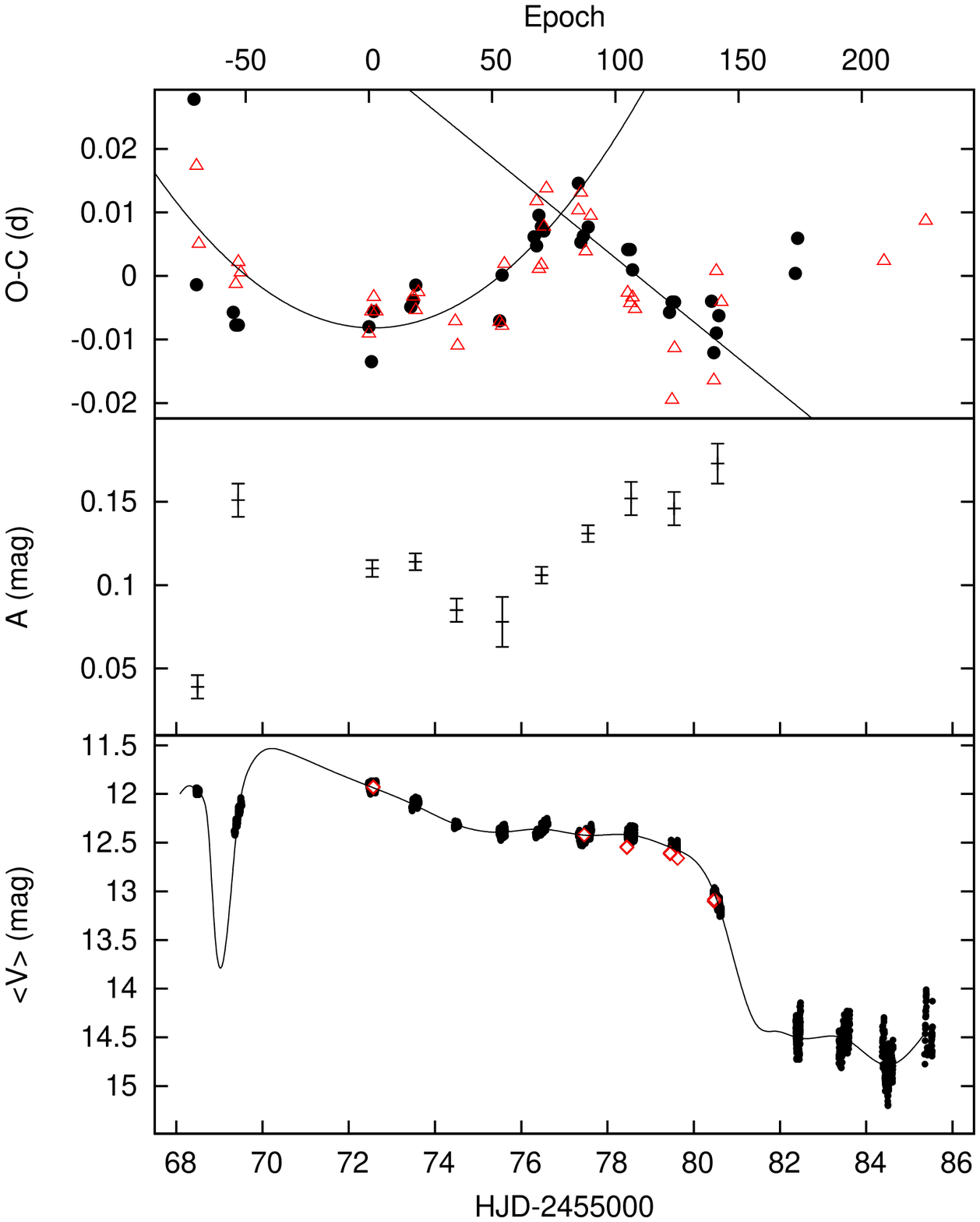}
\end{center}
\FigCap{
Top: The $O-C$ values corresponding to the particular phases of the superoutburst.
The black curve is the best fitting second order polynomial for cycles $E<80$.
The straight line is the best fit for cycles between $E=80$ and $E=150$.
Triangles and filled circles indicates $O-C$ values for maxima and minima 
respectively.
Middle: The maximum amplitude of the superhumps for each night.
Bottom: Our complete light curve of the 2009 superoutburst of 1RXS~J0532.
Diamonds denote the dates with $B,V,R,I$ measurements.
}
\end{figure}

\MakeTable{r | c c c c}{12.5cm}{$BVRI$ photometry of 1RXS~J0532 performed 
during the 2009 superoutburst.}
{
\hline \hline
Date \ \ \ \     &$B$          &$V$            &$R$        &  $I$  \\
\hline
28.08.2009 $\to$ &  $11.902\pm0.012$ &  $11.934 \pm 0.012$ &$11.039 \pm 0.011$  &$11.899 \pm 0.012$   \\
HJD-2455000 $\to$&  (72.569417)     & (72.569811)        & (72.570205)          &  (72.570598)        \\
                 &                  &                    &                      &                     \\
                 & $11.887\pm0.014$ &  $11.924\pm0.014$  & $11.014\pm0.016$     &  $11.881\pm0.012$   \\
                 &  (72.570992)     & (72.571386)        & (72.571779)          & (72.572173)         \\
\hline
02.09.2009 $\to$ & $12.411\pm0.011$ & $12.418\pm0.012$   & ---                  &  $12.33\pm0.012$    \\
HJD-2455000 $\to$& (77.454877)      & (77.457828)        & ---                  & (77.459414)         \\
                 &                  &                    &                      &                     \\
                 & $12.396\pm0.011$ & $12.42\pm0.011$    & ---                  & $12.323\pm0.012$    \\
                 & (77.455687)      & (77.458534)        & ---                  & (77.46012)          \\
\hline
03.09.2009 $\to$ & $12.537\pm0.013$ & $12.54\pm0.012$    & ---                  & $12.466\pm0.012$   \\
HJD-2455000 $\to$ &  (78.443584)     & (78.448480)        & ---                  & (78.451084)        \\
                 &                  &                    &                      &                    \\
                 & $12.538\pm0.012$ & $12.551\pm0.012$   & ---                  & $12.441\pm0.012$   \\
                 & (78.445910)      & (78.449846)        & ---                  & (78.452010)        \\
\hline
04.09.2009 $\to$ & $12.607\pm0.014$ & $12.602\pm0.013$   & ---                  & $12.495\pm0.014$   \\
HJD-2455000 $\to$ & (79.449179)      & (79.451066)        & ---                  & (79.453589)        \\
                 &                  &                    &                      &                    \\
                 & $12.605\pm0.012$  & $12.615\pm0.014$   & ---                  & ---                \\
                 & (79.450070)       & (79.452478)       & ---                   & ---                \\
                 &                   &                   &                       &                    \\
                 & $12.644\pm0.015$  & $12.660\pm0.014$   & ---                  & $12.534\pm0.015$   \\
                 & (79.626680)       & (79.627711)        & ---                  & (79.628463)        \\
                 &                   &                    &                      &                    \\
                 & $12.656\pm0.014$  & $12.660\pm0.014$   & ---                  & $12.557\pm0.016$  \\
                 & (79.631009)       & (79.628463)        & ---                  & (79.629215)       \\
\hline
05.09.2009 $\to$ & $13.099\pm0.013$  & $13.104\pm0.014$   & ---                  & $(12.895\pm0.014)$ \\
HJD-2455000 $\to$& (80.470725)       & (80.472692)        & ---                  & (80.474579)        \\
                 &                   &                    &                      &                    \\
                 & $13.117\pm0.013$  & $13.083\pm0.013$   & ---                  & $12.904\pm0.014$   \\
                 & (80.471639)       & (80.473676)        & ---                  & (80.475447)        \\
\hline
}
The lower panel of Figure 1 presents the photometric measurements of 1RXS J0532 
during our campaign.
In the complete light curve one can clearly see the precursor - 
a characteristic increase of brightness before the real superhump maximum. Such a precursor is occasionally observed 
for SU UMa stars, however, as far as we know there is no proven relationship
between the presence of a precursor and the orbital period, mass ratio or
other parameter of the system.

There was only one night in our campaign (24 August) that allowed us to analyse the
characteristics of the precursor. In the light curve from this night, a quasi sinusoidal modulation
can be observed with average brightness of $V=11.97$. A rough peak-to-peak distance
measurement gives $P_{p}=0.059$d with an uncertainty of the order of $0.005$d. This
uncertainty is too large to draw conclusions about early superhumps or compare $P_{p}$ with the orbital period. Nevertheless,
the presence of the precursor is certain.

During the observation on 25 August, a sharp rise in brightness was
detected. However the initial level was $V=12.45$,
about 0.5 magnitude less than during the previous night. 
By this time the precursor stage had ended and the brightness of the star
was increasing towards the real superoutburst maximum.

Similar behaviour of the overall superoutburst light curve 
of 1RXS J0532, especially the presence of the precursor, has been reported
by Imada et al. (2009). They studied the 2005 superoutburst in terms of 
the TTI model. Here, we propose a different interpretation of this phenomenon.
We consider the model in which a superoutburst is
caused by enhanced mass transfer from the secondary (Smak 2008ab; 2009b).
A rapid decline in the brightness at the end of the precursor is most likely
an indication that the normal outburst is ending.

However, the disc
at this moment is already heated, so irradiation of the secondary by the disc is intense.
Due to the irradiation, the secondary star is heated and the mass transfer increases again. This
consequently leads to an increase in the brightness of the hot spot and the 
whole system. 

In addition, oscillations of the hot spot (due to variable dissipation
of kinetic energy of the stream) could be responsible for superhumps
observed during the precursor and the rest of the superoutburst.

We do not observe a classic plateau after the maximum, in contrast to the observations of Imada et al. (2009).
Instead we see a clear deviation from a straight line
with a prominent change in slope around HJD 2455075, a clear indication that the rate of
mass transfer can vary.


After 5 September, a fast decrease in brightness was detected. At
minimum, the observed star was observed at around $V=14.8$.

Unfortunately, due to observing schedule constraints, we were not able 
to collect data on 26 and 27 August. Therefore 
we have to estimate that at maximum the superoutburst
reached a brightness close to $V=11.5$ and that the superoutburst amplitude was around
$V=3.3$.

\subsection{Color variation}
Most of the data were gathered without filters but several exposures were
taken in $B,V,R$ and $I$ filters. Table 2. presents the results from these
measurements. Two comparison stars, 2MASS~J05323636+6245536 and 2MASS~J05325096+6246161, were used to derive these results.
The measured $B-V$ colour indices (in the -0.34 to 0.19 range) correspond to the known colour index
for dwarf novae in outburst (e.g. Bailey 1980, Echevarria \& Jones 1983).
While it would have been desirable to measure the colour index of the early superoutburst,
this was not possible due to observing constraints.

\Section{Superhumps}
\begin{figure}[!h]
\includegraphics[height=0.9\textheight,width=\textwidth]{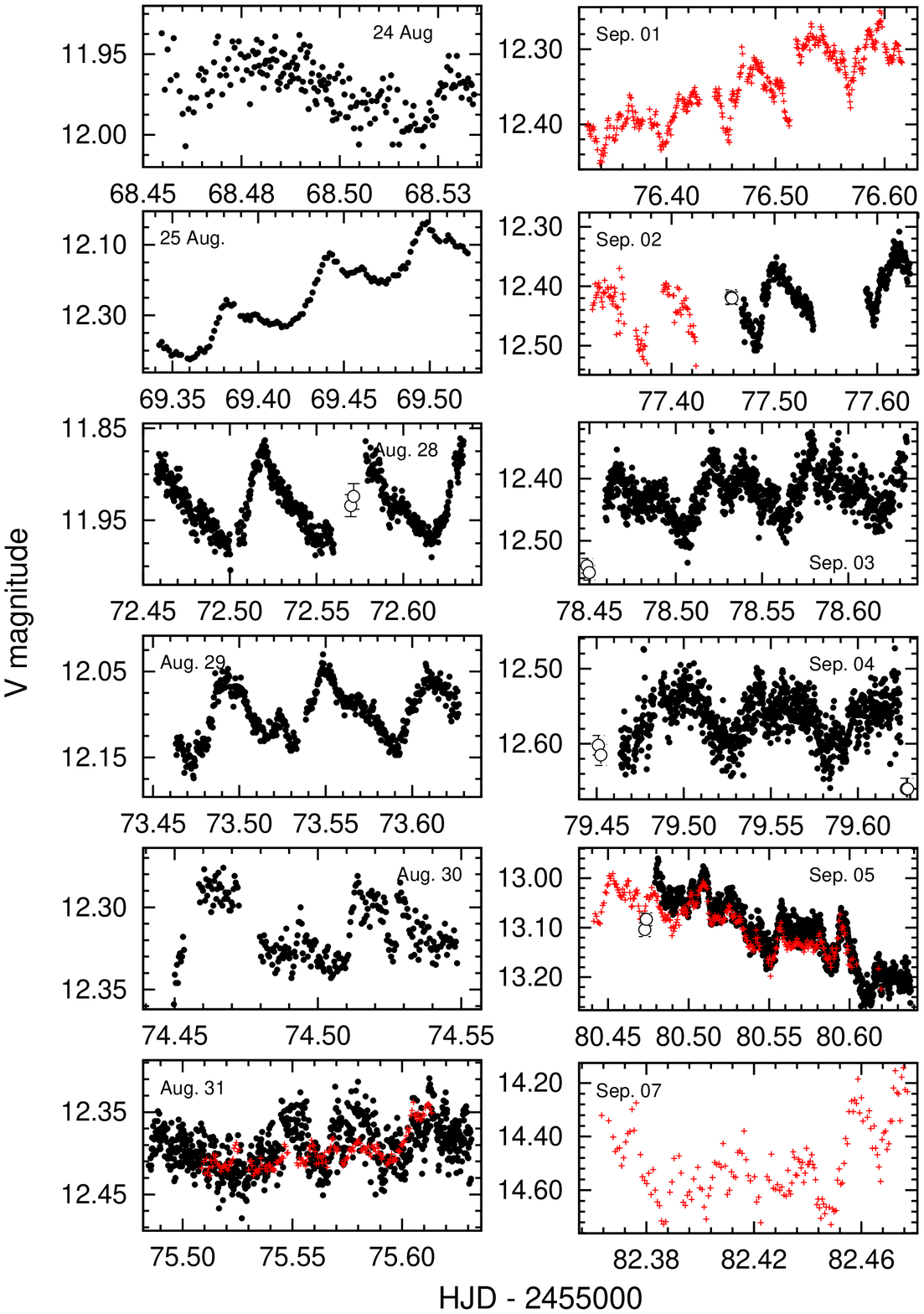}
\FigCap{Evolution of superhumps in the light curve of 1RXS J0532 during the 2009 superoutburst. Filled circles
correspond to the observations performed at Skinakas Observatory, except on Aug 24 and Aug 25. Crosses denote data
from the Ostrowik Station. Open circles show measurements taken in the Skinakas Observatory with the $V$ filter, used during the calibration procedure.
}
\end{figure}
Superhump evolution during our observing campaign can be followed in Figure 2.
At the beginning of our observing run on 24 August only a hint of modulations was visible.
However the second night revealed the development of superhumps on the rising branch of the superoutburst with 
an amplitude of ~0.15 mag.
Fully formed and regular superhumps were visible on 28 August but with a lower amplitude
of approximately 0.11 mag. This is a consequence
of the fact that the superoutburst maximum occurred most probably two days before 28 August. However, the superhump amplitude was 
relatively small, especially if we assume the standard $dA/dt\sim 0.015$ (mag/day)
and compare it with an amplitude of $\sim 0.28$ (mag)
at the superoutburst maximum read from figures in Imada et al. (2009) and
Poyner \& Shears (2006).

The evolution of superhumps after 28 August is also very 
interesting. Firstly, signs of secondary humps can be 
noticed on 29 and 30 August. Secondly, after 31 August the minima became much
more prominent and clearer than
the rather fuzzy maxima, in contrast to the situation before 31 August when 
the maxima were mostly much sharper than the minima.
Moreover, the minima of the light curve on 1 September suggest that
the bright spot became fuzzy and reduced in size. The superhump source was in the
most part hidden behind the accretion disc, but only for a very
short time as manifested by the sharp, narrow dips in the light curve
during that night.
Such hot spot radius variations are in agreement with the hot spot oscillation
model and hot spot eclipses observed in OY Car or Z Cha (Smak 2007,2008b).
Why did such transitions occur around 1 September and not before? 
To answer this question we refer to Smak (2010), Figure 2. The main conclusion
drawn from that figure is that the superhump source is located close to but
above the disc rim. According to the work of Smak(2007, 2008ab), we 
assume that the mass
transfer rate can vary during a superoutburst. This allows us to conclude
that the distance of the superhump source from the disc plane may be changing as well.
In particular,
the source is going to be more fuzzy and closer to the disc plane as the 
superoutburst declines.

After 5 September the system started fading quickly. Strong
flickering became very prominent. Starting on 7 September, the flickering very effectively masked other periodicities and only hints
of superhump modulation can be seen.
\subsection{$O-C$ analysis}
During the whole superoutburst the light variation of 1RXS~J0532 shows clear
but complex modulations (Figure 2).
Previous studies have found a similar behaviour in other dwarf 
novae systems. In some cases even alternating superhump periods
have been reported
(eg. Olech et al. 2004, Rutkowski et al. 2007). In order to study this behaviour in detail we investigated 
the maxima and minima present in our light curve of 1RXS~J0532 adopting the following procedure.

\MakeTable{c c c c | c c c c}{12.5cm}{Times of the extrema in the lightcurve of
1RXS~J0532 during the 2009 superoutburst.}
{
\hline 
E    & HJD* max   & Error     & $O-C$      & E   & HJD* min  & Error      & $O-C$ \\ \hline
-70  & 68.4850 & 0.0071 & 0.017303 & -71 & 68.4819 &    0.0041  & 0.027830 \\
-69  & 68.5299 & 0.0030 & 0.005040	& -70 & 68.5098 & 0.0021  & -0.001368 \\
-54  & 69.3807 & 0.0010 & -0.001273& -55 &69.3619  &    0.0007  & -0.005732 \\
-53  & 69.4413 & 0.0021 & 0.002194	&-54  & 69.4170 & 0.0007  & -0.007730 \\
-52  & 69.4968 & 0.0011 & 0.000541	& -53 & 69.4741 & 0.0011 & -0.007727 \\
0   &  72.4586 & 0.0020 & -0.009080& 0   &  72.5000&    0.0008 & -0.008000 \\
1   & 72.5193  & 0.0006  & -0.005593& 1  &  72.5516 &   0.0005 & -0.013498 \\
2   & 72.5787  & 0.0004  &-0.003306&  2  &  72.6166 &   0.0002 & -0.005595 \\
3   & 72.6336  & 0.0007  &-0.005579& 17  & 73.4738 &    0.0008 & -0.004859 \\
18  & 73.4930  & 0.0006  &-0.003272& 18  & 73.5319 &    0.0005 & -0.003857 \\
19  &73.5480  & 0.0007  &-0.005395& 19  & 73.5914 &    0.0002 & -0.001454 \\
20  &73.6080  & 0.0007 &-0.002548  &53  &75.5271  &    0.0004 & -0.007073 \\
35  & 74.4606 & 0.0021 & -0.007131	&54  & 75.5914 &  0.0011 & 0.000130 \\
36  & 74.5139 & 0.0017 &-0.010994	&67  & 76.3397 &  0.0005 & 0.006161 \\
53  & 75.4890 & 0.0009 & -0.007274	& 68 & 76.3954 &  0.0004 & 0.004763 \\
54  & 75.5456 & 0.0011 & -0.007847	& 69 & 76.4573 &  0.0005 & 0.009566 \\
55  & 75.6125 & 0.0012 & 0.001891	& 70 & 76.5126 &  0.0004 & 0.007768 \\
68  & 76.3652 & 0.0008 & 0.011753	& 71 &  76.5690&  0.0005 & 0.007070 \\
69  & 76.4116 & 0.0007 & 0.001080	& 85 &  77.3759&  0.0004 & 0.014604 \\
70  & 76.4694 & 0.0006 & 0.001737	& 86 &  77.4237&  0.0008 & 0.005306 \\
71  & 76.5326 & 0.0008 &0.007744	& 87 & 77.4818 &  0.0008 & 0.006309 \\
72  & 76.5957 & 0.0007 & 0.013751	& 89 &  77.5974&  0.0005 & 0.007714 \\
85  &77.3352  & 0.0008 & 0.010313	& 105& 78.5074 &  0.0009 & 0.004152 \\
86  & 77.3951 & 0.0006 & 0.013111	& 106& 78.5645 &  0.0010 & 0.004154 \\
88  & 77.5002 & 0.0005 & 0.003875	& 107& 78.6184 &  0.0007 & 0.000957 \\
90  & 77.6200 & 0.0006 & 0.009469	& 122& 79.4682 &  0.0007 & -0.005707 \\
105 & 78.4650 & 0.0012 & -0.002664	& 123& 79.5269 &  0.0008  &-0.004105 \\
106 & 78.5205 & 0.0011 & -0.004307	& 124& 79.5840 &  0.0008 &-0.004102 \\
107 & 78.5786 & 0.0015 & -0.003390	& 139& 80.4406 &  0.0004 & -0.003966 \\
108 & 78.6339 & 0.0017 & -0.005213	& 140& 80.4896 &  0.0011 &-0.012064 \\
123 & 79.4768 & 0.0018& -0.019517	& 141& 80.5498 &  0.0008 &-0.008962 \\
124 & 79.5421 & 0.0013 & -0.011370	& 142 & 80.6096 & 0.0005 & -0.006259 \\
140 & 80.4513 & 0.0010 & -0.016456	& 173 & 82.3863 & 0.0006 & 0.000415 \\
141 & 80.5256 & 0.0009 & 0.000771	& 174 & 82.4489 & 0.0014 & 0.005918 \\
143 & 80.6350 & 0.0017 & -0.004105	& & & & \\
209 & 84.4129 & 0.0012 & 0.002344	& & & & \\
226 & 85.3907 & 0.0013 & 0.008655	& & & & \\
\hline
* HJD - 2455000
}
Firstly, we de-trended the global light curve. For each night's observations we subtracted a first or second order 
polynomial and derived a periodogram. The ZUZA code (Schwarzenberg-Czerny 1996) and its {\tt perort} procedure was
used to obtain the power spectrum. The strongest signal is at the period
$P=0.05714\pm0.00010$d ($f=17.4994$c/d, Figure 3).
In order to check if the derived period is stable we constructed an $O-C$ diagram.
We determined 37 times of maxima and 34 times of minima (Table 3). Linear regression gave
\begin{equation}
HJD_{\rm max} =  72.4677(19) + 0^{\rm d}.057143(19) E,
\end{equation}
for the maxima, and
\begin{equation}
HJD_{\rm min} = 72.508(2) + 0^{\rm d}.057098(21) E,
\end{equation}
for the minima, where $E$ is the cycle count (Epoch).
Equations 1 and 2, for maxima and minima respectively,
have been used to construct the $O-C$ diagram presented in the upper panel of 
Figure 1.
The times of extrema can be fitted with a quadratic
ephemeris (for $E<80$) which are given by the following equations 

for maxima and minima respectively:
\begin{eqnarray}
\begin{array}{lllll}
HJD_{\rm max} = & 72.4673 & +0^{\rm d}.0571117E & + 3.05\cdot10^{-6}E^{2} \\
                &\pm0.002 &\pm0.000022& \pm 0.55 
\end{array}
\end{eqnarray}
and
\begin{eqnarray}
\begin{array}{lllll}
HJD_{\rm min} =& 72.4673 & + 0^{\rm d}.0571069E& + 3.19\cdot10^{-6}E^{2}  \\
               &\pm0.0029 &\pm0.0000344        & \pm 0.89  
\end{array}.
\end{eqnarray}
The period derivatives calculated from Eqs. (3) and (4) separately for
maxima and minima differ slightly ($\dot{P}_{\rm max}=(10.68 \pm 1.9)\cdot 10^{-5}$ and
$\dot{P}_{\rm min}=(11.18 \pm 3.1)\cdot 10^{-5}$). Thus we decided to derive
the rate of period change based on an $O-C$ analysis for the maxima and
minima together.

For $E<80$ we fitted a second order polynomial.
The best-fitting quadratic equation was:
\begin{equation}
O-C(days) = - 7.88(1.29) \cdot 10^{-3} + 1.17(1.68) \cdot 10^{-5}E + 2.64(41) \cdot 10^{-6}E^{2}.
\end{equation}
Equations (1,2,5) can be used to determine the $P_{sh}$ derivative. From the definition we have:
\[
O-C = a + b E + \underbrace{\frac{1}{2}\frac{dP}{dt}\bar{P}}_q E^{2} 
\]
so
\[ \dot{P} = \frac{dP}{dt}=\frac{2 q}{\bar{P}},\]
where $\bar{P}$ is the average period used to generate the $O-C$ diagram.
We take the weighted mean of the periods from Eq. 1 and 2 to obtain $\bar{P}$ = $P_{sh}$ = 0.057122(14)d.
Thus for parameters from Eq. 5 we get:
\begin{equation}
\dot{P} = (9.24 \pm 1.4) \cdot 10^{-5} \ {\rm or}  \ (5.28 \pm 0.04)\cdot 10^{-6} (days/cycle).
\end{equation}
This is the rate at which the superhump period ($P_{sh}$) increased during
the so-called stage B (see paragraph 3.2 in Kato et al. 2009). 
We fitted a constant function to the cycles between 80 and 150 and obtained
\begin{equation}
O-C_{\rm 80-150}(days) = -0.000316(51) E + 0.0344(58),
\end{equation}
where the coefficient before E suggests a constant period equal to
$P_{q}=0.056806(51)$d (that is shorter than $P_{sh}$ by approximately
$0.000316$d).
It is worth noting that Kato et al. (2010) report that period analysis 
for 1RXS J0532 of their data
collected during quiescence reveals a period of $0.5690(2)$d.
Most likely the variation found by us has the same physical nature 
as the modulation found by Kato et al.

Extrema detected
for $E>150$ do not follow the same constant trend and probably come from
orbital modulations of the hot spot.
\Section{Power spectrum analysis}
\begin{figure}[htb]
\includegraphics[width=\textwidth]{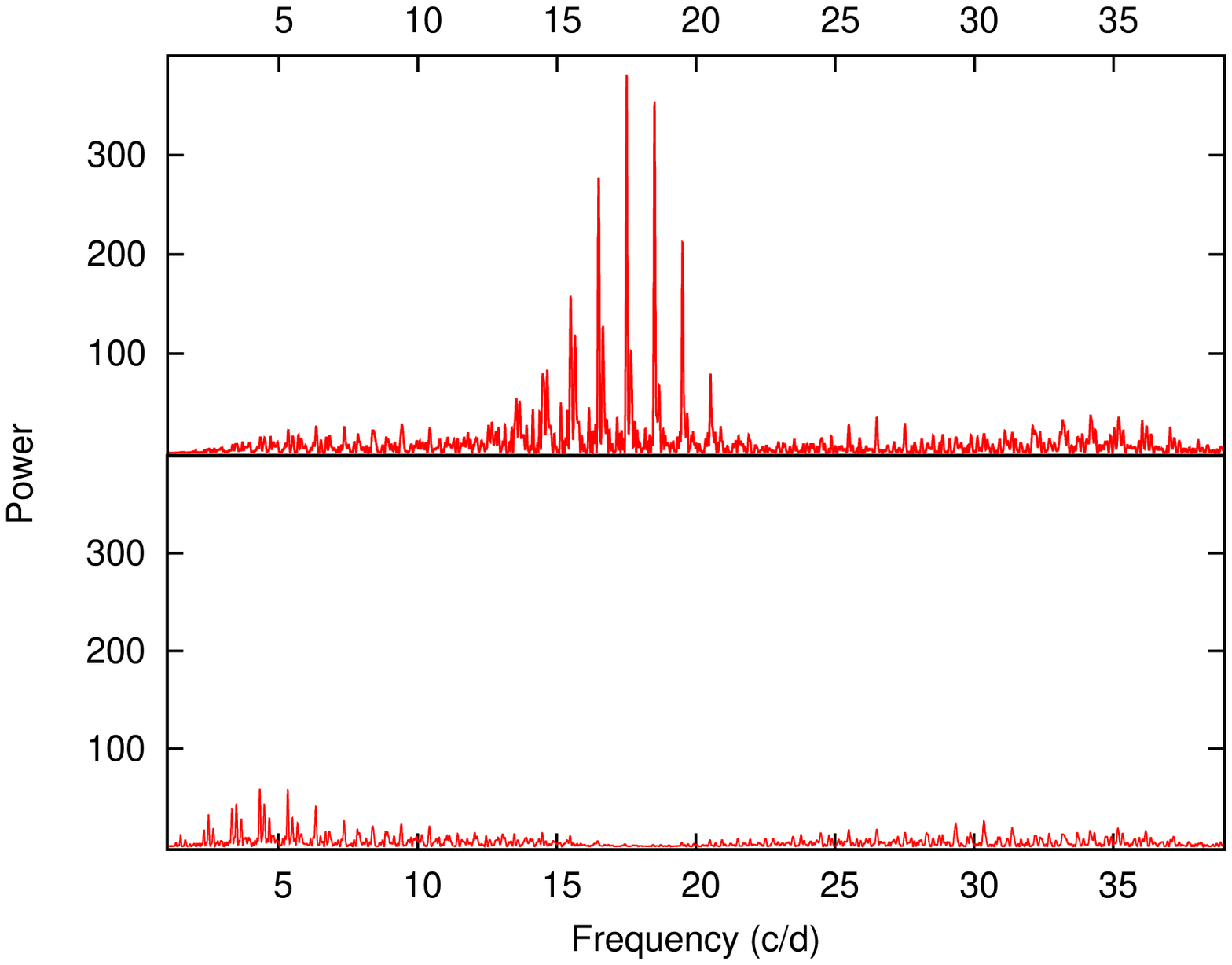}
\FigCap{
Top: Power spectrum of 1RXS J0532 for the complete observing run during the 2009
superoutburst. The most significant peak occurs at frequency $f_{sh}=17.4994\pm0.0323$c/d
corresponding to period $P_{sh}=0.05715(10)$d. Bottom:
Pre-whitened spectrum after removing the dominant frequency corresponding to the above period. The most
prominent reamining peaks are located at frequencies $f=4.329$c/d and $f=5.334$c/d.
} 
\end{figure}

Figure 3 shows the power spectrum of 1RXS J0532 obtained for all the data collected
during the 2009 superoutburst. The Analysis of Variance (AOV) method with
3 harmonics was used to analyse it.
The strongest peak is at frequency $f=17.4994\pm0.0323$c/d, which 
corresponds to a period of $P=0.057145(105)$d, consistent with the superhump period of the system.
However the error is about one order of magnitude larger than that obtained
through the $O-C$ analysis. Therefore we finally adopt the value derived
from the $O-C$ analysis, $P_{sh}=0.057122(14)$d, as our most accurate measure of the superhump period.

Superoutburst light curves of some dwarf novae contain additional 
periodicity which can be revealed by the pre-whitening procedure. Such light
modulations are often connected to the orbital period or other kind of superhumps (eg. Olech et al. 2009, Rutkowski et al. 2009).
We checked if such a periodicity can also be detected in the case of 1RXS J0532. In order to do that, from the 
de-trended light curve of a particular night,
we individually subtracted the frequency
$17.4994$c/d (corresponding to the superhump period).
We fitted a Fourier series by the least squares method and subtracted 
it from the data. Next, all light curves obtained in this way were combined 
and a power spectrum computed.
This method of pre-whitening allowed us to remove the influence
of variable amplitude and light curve shape on the spectrum. The periodogram
obtained with this method is presented in the bottom panel of Figure 3.
Lets assume that some other periodicity (like orbital period or negative
superhumps period) close to $P_{sh}$ was present before in the light curve.
Those variations should be clearly visible in the pre-whitened spectrum
(like for instance in Olech et al. 2007).
However, there is no significant signal around $f\sim17.5$c/d distinguishable
from the noise and only a low-power multiplet is visible. Two highest peaks
in this multiplet at frequencies $f_{1}=4.329\pm0.036$ and $f_{2}=5.334\pm0.036$ 
are visible. The nature of those peaks is unknown. This could be the 
effect of imperfect subtraction of the signal $P_{sh}$ or its alias.
However, it could also be a result of physical processes going on
in this system, especially because other systems also show peaks in
the same frequency range. In particular observations of WX Hyi and 
SDSS J162520.29+120308.7 
(Olech et al. 2011 in preparation, private communication) reveal
peaks at frequencies around 5-6 c/d in the quiescence spectrum.

\subsection{High-frequency variations}
Relatively large scatter in the light curves observed on 30 and 31 August and 3 and 4
september suggest the presence of additional variability.
We expect that the oscillations responsible
for this scattering must be much shorter than $P_{sh}$.
Dwarf nova oscillations or quasi-periodic oscillations
(QPO) can be a reasonable explanation for such modulations (eg. Woudt \& Warner (2002), Woudt et al. (2010) and references 
therein). We checked if such oscillations were present in our 
light curve.

The observing run on 30 August is too short to look for quasi-periodic 
variability. However, observations performed on 31 August, 3 and 4 September consist of long runs, with exposure times around 10s. The data have
good quality and the variable star has a signal level $\sim8000$ADU over the noise which results
in errors of the order of 0.01 mag. Hence, we looked for QPO variability in the
data from these three nights.

We applied the following procedure to the data from 31 August, 3 and 4 September.
First we removed the global superoutburst trend and then,
for each night separately, we removed the modulations responsible
for the superhumps by fitting a Fourier series with 10 harmonics. We then
calculated the power spectra.

The complex character of the superhump light curve prevents a complete removal of
all frequencies connected with the superhumps.
We only present results for 3 September for which the above procedure gave the 
best result ie. the most prominent peaks are higher than 4$\sigma$
(over $99.99\%$ confidence level).
Figure 4 shows the AOV periodogram obtained with 2 harmonics.
We show the spectrum from 160 c/d as low frequency peaks don't contain any 
relevant information but only information about less than perfect removal of 
the non-strictly-periodic superhump modulation and noise.
\begin{figure}[htb]
\includegraphics[angle=-90, width=\textwidth]{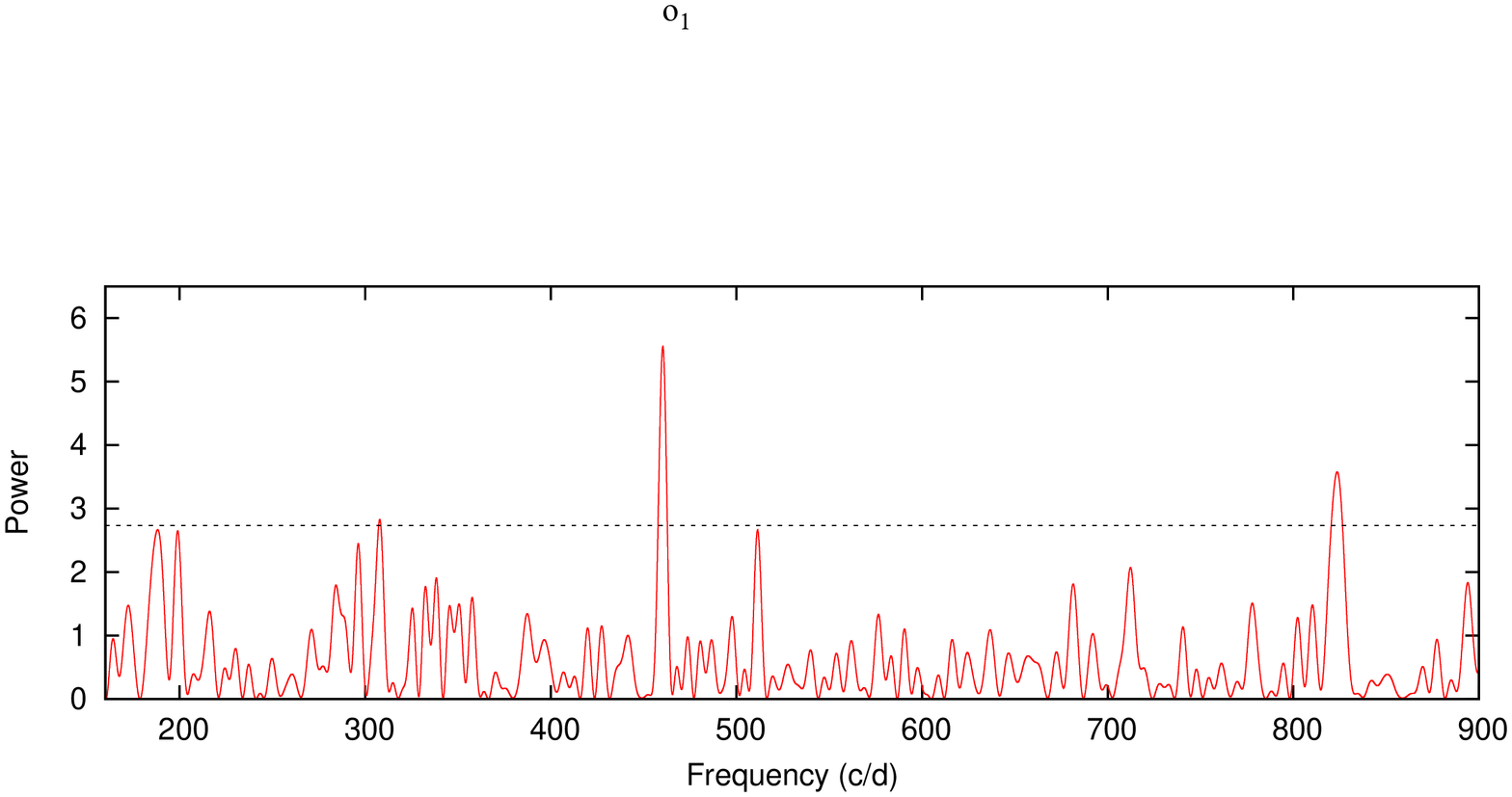}
\FigCap{The AOV periodogram obtained for 3 September. The two highest peaks are
 located at frequencies
$f_{1}=460.34\pm 1.62$c/d and $f_{2}=823.96\pm3.35$c/d. The dotted horizontal line
represents the 4$\sigma$ confidence level.
}
\end{figure}

Two particularly clear peaks are present in the periodogram at 
frequencies
$f_{1}=460.34\pm1.63$c/d (188s) and $f_2=823.96\pm3.35$c/d (104s).
The corresponding periods, and their low coherence, are characteristic
of the quasi periodic oscillations observed in some dwarf novae.
Although the spectrum we obtained looks credible, the high humidity that occurred during these nights could have
caused condensation on the glass covered CCDs and affected our measurements
and greatly increased the noise.
\subsection{Apsidal motion of accretion disc}
Smak (2009a) presented several arguments that the evidence presented earlier
by several authors for 
disc eccentricity in dwarf novae is a result of measurement errors or 
arbitrary incorrect assumptions. However he did not rule out the existence of an eccentric accretion disc.
In the TTI model 
such eccentricity is a crucial ingredient and thus its absence would result in the
rejection of the entire TTI  model. But even for the Smak interpretation of 
superhumps, the presence of such an eccentric disc which "precesses" 
(in the sense of apsidal motion) would give the necessary clock for superhumps
timing.

From Figure 1 one might expect a decreasing trend of 
brightness from night to night consistent with the globally
decreasing trend of the plateau region. Instead, on some nights we clearly see
a trend of increasing brightness. 
One might suspect that atmospheric extinction could be responsible for this
behaviour. However, since we performed differential photometry, only the
second order extinction coefficient $k''$ could cause this. We determined the
colour difference between the variable and comparison stars to be about 0.3
mag. Assuming quite a high, but still reasonable, value of $k''$, we find that
for airmass between 1 and 2 the change in brightness due to differential colour
extinction cannot be larger than 0.02 mag. This is about one order of
magnitude too small to produce the observed behaviour. Moreover, we studied
over a dozen stars in the field and only 1RXS J0532 exhibited this trend.
Even if we assume that this is an instrumental effect, it cannot explain the
observed brightness oscillation around the mean trend of the plateau.

This raised our suspicion that an additional periodicity
may be present in the light curve. 
\begin{figure}[htb]
\includegraphics[angle=-90,width=\textwidth]{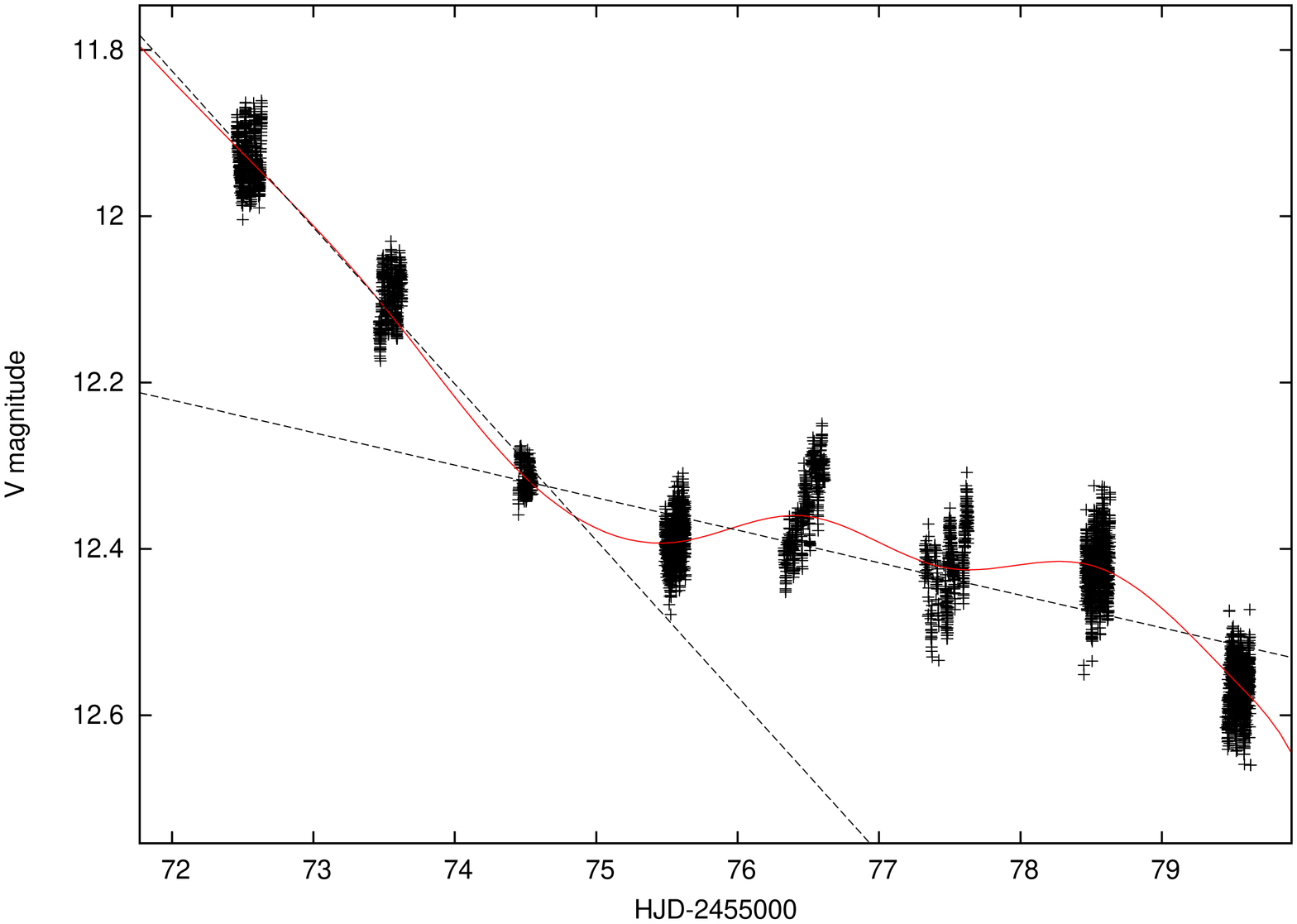}
\FigCap{A magnification of the 2009 superoutburst plateau. Dashed straight lines
represent the best least squares linear fits to the data taken on 28 - 30 August and
31 August - 4 September. The solid curve indicates the observed oscillation around the trend
represented by the dashed lines.
}
\end{figure}

To investigate this behaviour further, we fitted two linear functions to model the plateau stage of
the superoutburst. Two functions were needed because of the shape of the observed plateau
(see section 3). The dashed lines in Figure 5 represent these linear fits to 
the corresponding parts of the superoutburst. For each night shown in Figure 5 the mean values of time and magnitude were derived.
An indicative cubic spline curve through these mean values is shown to highlight the
variations in brightness around the straight lines.

In order to measure the period of this modulation, for each night shown on Figure 5 we calculated average values \footnote{according to
$(sh(m,t)_{\rm max_1}+sh(m,t)_{\rm min_1})/2, (sh(m,t)_{\rm min_1}+sh(m,t)_{\rm max_2})/2, (sh(m,t)_{\rm max_2}+sh(m,t)_{\rm min_2})/2 \ldots$}
 for successive pairs of maxima and minima.
In doing this we have assumed that each extremum is defined by two parameters:
brightness - $m$ and time - $t$.
Finally, we subtracted trends modeled previously with the linear functions. The resulting residuals are presented in Table 4. 

\MakeTable{c c | c c | c c}{12.5cm}{Residual values of mean magnitude with respect to linear fits to the lightcurve of
1RXS~J0532 during the 2009 superoutburst.}
{
\hline
HJD* (d)    & res (mag)  & HJD* (d)  &  res (mag)  & HJD   & res(mag) \\ \hline
72.4793 &  0.01726 & 75.5600 &  0.03376& 77.4098  &  0.01363 \\
72.5099 &  0.00436 & 75.5720 &  0.02074& 77.4913  & -0.00465 \\
72.5367 & -0.00443 & 75.5903 &  0.01157& 77.5211  & -0.02241 \\
72.5662 & -0.00954 & 75.6067 &  0.00948& 77.6096  & -0.05293 \\
72.5978 & -0.01596 & 76.3331 &  0.03623& 78.4862  & -0.03166 \\
72.6252 & -0.02317 & 76.3501 &  0.01641& 78.5148  & -0.04767 \\
73.4826 & 0.004494 & 76.3782 &  0.00607& 78.543   & -0.06373 \\
73.5123 & -0.01504 & 76.4106 & -0.00069& 78.5712  & -0.06583 \\
73.5399 & -0.02862 & 76.4411 & -0.00699& 78.5991  & -0.07422 \\
73.5704 & -0.03136 & 76.4684 & -0.02820& 78.6264  & -0.07129 \\
73.5999 & -0.03157 & 76.4961 & -0.04059& 79.4857  &  0.05210 \\
74.4569 &  0.02922 & 76.5227 & -0.06353& 79.5161  &  0.05341 \\
74.4843 &  0.01618 & 76.5510 & -0.07693& 79.5345  &  0.05609 \\
74.5093 &  0.01461 & 76.5827 & -0.08707& 79.5632  &  0.06177 \\
75.5086 &  0.04597 & 77.3510 &  0.03033& 79.6042  &  0.06172 \\
75.5426 & 0.039046 & 77.3845 &  0.03047&          &          \\
\hline
* HJD - 2455000
}
For the residuals we calculated the AOV periodogram with five harmonics 
using the ZUZA code. The results are 
presented in Figure 6. The most prominent peak is located at
$f=0.5885\pm0.014$c/d which corresponds to $P=1.6992$ days.

\begin{figure}[htb]
\includegraphics[angle=-90,width=\textwidth]{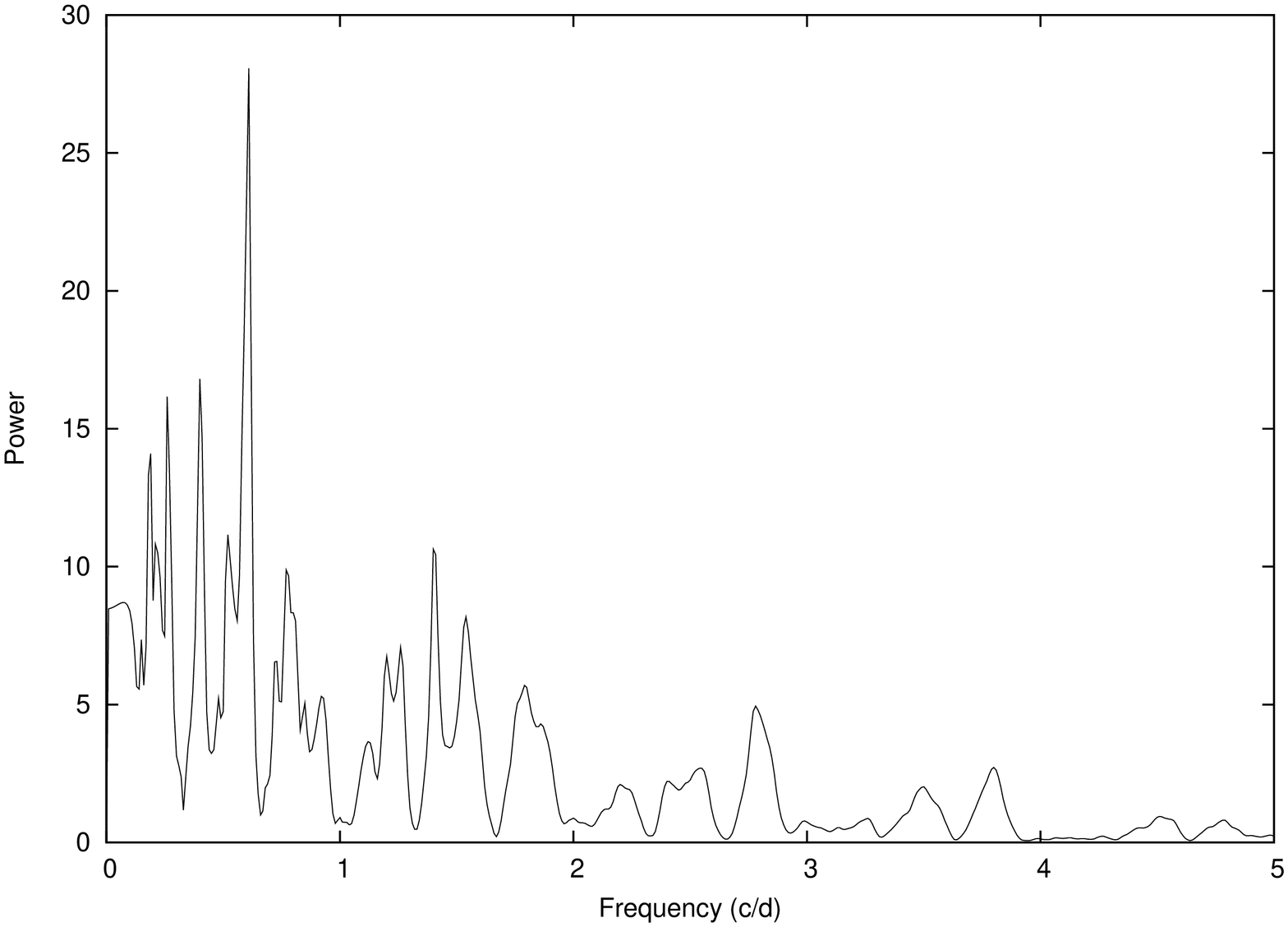}
\FigCap{Power spectrum obtained for residuals calculated after subtracting
the linear trend of the plateau from the mean superhump magnitudes. The highest
peak is located at frequency $f=0.5885$c/d.  
}
\end{figure}
\begin{figure}[htb]
\includegraphics[angle=-90,width=\textwidth]{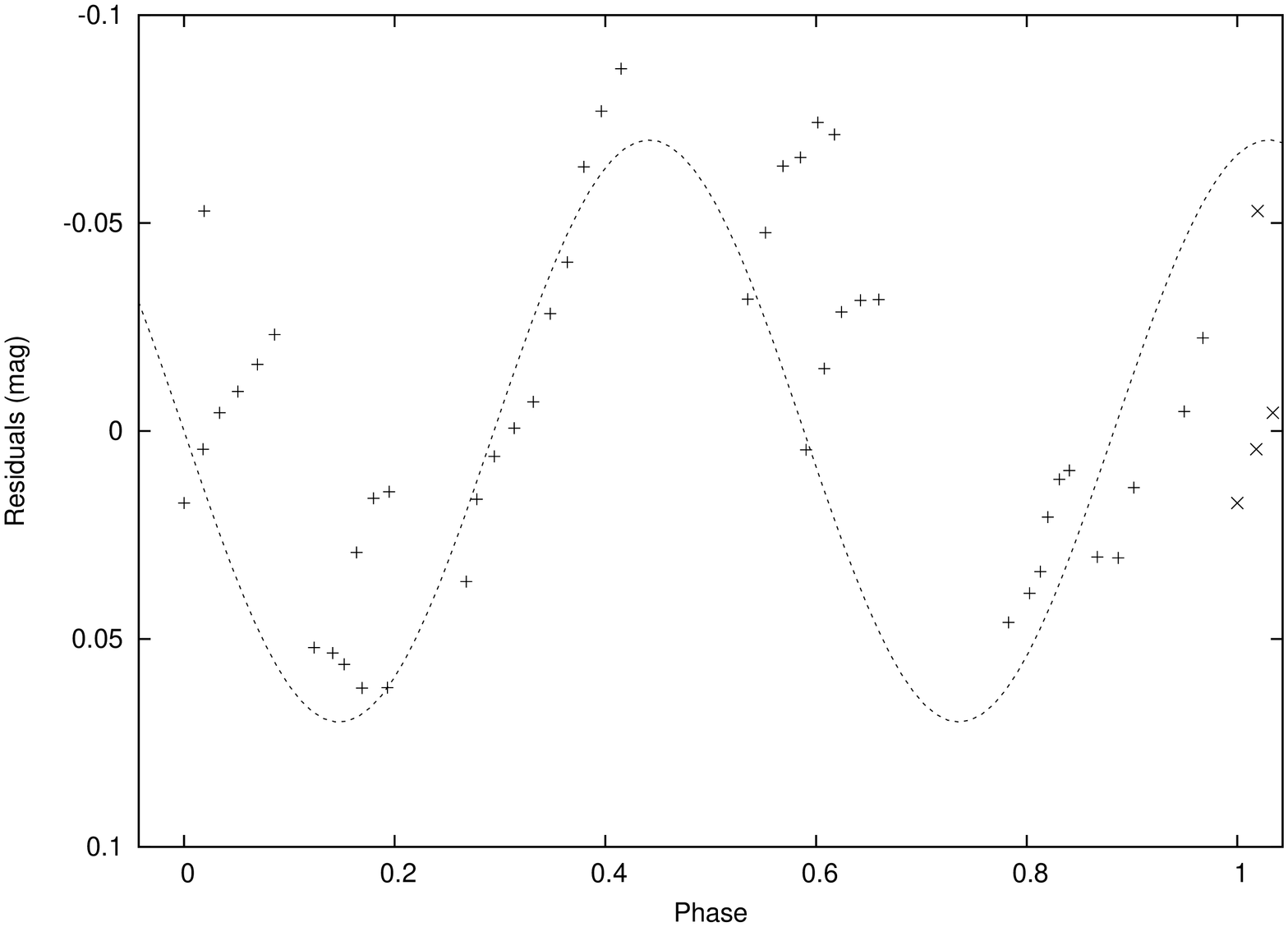}
\FigCap{Phased light curve of residuals with period $P=1.6992$d 
obtained after subtracting the linear trend. The solid line shows
a sine function with the above period.
}
\end{figure}

Figure 7 presents a residual light curve phased with this period. 
In addition a sine function with this period is superimposed on the residuals. Although the scatter is substantial,
the periodic modulation can be clearly seen.

There is a well know relation between the beat period $P_{prec}$, the orbital period $P_{orb}$ and the
superhump period $P_{sh}$ observed during a superoutburst:
\begin{equation}
\frac{1}{P_{prec}} = \frac{1}{P_{orb}} - \frac{1}{P_{sh}}
\end{equation}
For $P_{sh}=0.05712$d and $P_{orb}=0.05620$d we obtain $P_{prec}=3.39$d.

If the narrower side of the eccentric accretion disc is turned towards the observer 
we observe minimum brightness. When the more extended side of the disc is turn toward
the observer we detect maximum brightness. Hence two maxima and two minima caused by
the orientation of the accretion disc should be visible per orbital period.

Since we measured a modulation with period $P=1.6992$d, we interpret this to be the effect of apsidal
motion of the accretion disc with period $P=P_{prec}/2$.

Such apsidal oscillations should mostly be visible in objects with 
high orbital inclination, which is in agreement with the high $i$ value measured for 1RXS J0532.
\Section{Conclusions}
\begin{itemize}
\item[a.] It has been confirmed that 1RXS J053234+624755 is a SU UMa star. 
The measured superhump period is $P_{sh}=0.057122(14)$d. The superoutburst 
is characterized by a slight rebrightening in the later phase of the plateau.
The amplitude of the superoutburst was determined to be 3.3 mag. 
\item[b.] 
We determined the rate of superhump period change to be $\dot{P}=9.5 
\cdot 10^{-5}$. This value is noticeably different than the rates obtained for
for superoutbursts in 2005 and
2008 ($\dot{P}$=$5.7 \cdot 10^{-5}$ and $10.2 \cdot 10^{-5}$, Imada et al. 2009 and Kato et al. 2009, respectively).
%
%
\item[c.] We proposed that the observed superhump light curve behavior
can be explained if we assume that the distance of the superhump source from
the disc plane decreases as the superoutburst declines.
\item[d.] Additional data analysis allowed us to 
detect a quasi-periodic signal at the frequency $f_1=460.34\pm1.63$c/d
(188s). This frequency, however, may be spurious and simply the result
of high humidity during the observation.
\item[e.]
Detailed analysis of the superoutburst plateau phase enabled us to detect
oscillations with a period $P_{prec}/2=1.699\pm0.005$d, which we interpret as the effect
of an apsidal motion of the accretion disc.
\item[f.] Based on the double peaks in the emission lines,
Kapusta \& Thorstensen (2006) concluded that the orbital inclination, $i$,
of 1RXS J0532 is not far from edge on. Given the prominence in the spectroscopic data 
of the double peaks in the emission lines, it is almost certain
that $i$ is larger than $70$ deg. Taking into account the fact that
there are no observed disc eclipses, we have assumed $i=75$ deg.

Even though we were not able to observe this system during the maximum of the
superoutburst, visual inspection of the
amplitude of superhumps in Figure 4 in Poyner \& Shears (2006)
and Figure 3 in Imada et al. (2009) gave $A_{0}=0.28\pm0.01$ mag.
Using eq. (4) and (5) from Smak (2010) we obtained $A_{n}=0.151$ mag.
These results are consistent with the dependence between superhump amplitude and
orbital inclination discovered by Smak (2010).
\end{itemize}
\Acknow{This work was partly supported by the Polish MNiSW grant
no. N203~301~335. Artur Rutkowski has been supported by 2221-Visiting Scientist
Fellowship Program of TUBITAK. AR is also grateful to Prof. Zeki Eker for
inspiring discussions. Skinakas Observatory is a collaborative project of 
the University of Crete and the Foundation for Research and Technology-Hellas.
We would like to thank the Skinakas Observatory staff for their help 
and assistance.
}

\end{document}